\documentstyle[epsf,epsfig,aps]{revtex}
\begin{document}

\pagestyle{myheadings}


\title{Bistable analytic-phase synchronization in strongly competing \\
chaotic lasing modes.
}

\author{
Sebastian Wieczorek and Weng W. Chow
}

\address{
 Sandia National Laboratories, Albuquerque NM 87185/0601, USA
 }

\date{\today}

\maketitle

\begin{abstract}
                                                                       
We theoretically study  analytic-phase synchronization in strongly-competing 
oscillator systems. Using the example of composite-cavity modes coupled via
a class-B laser active medium, we discover that 
inherent chaotic phase 
synchronization can  arise concurrently at two different chaotic attractors,
leading to bistable phase-synchronized solutions.
In our example, the underlying mechanism for  bistability and 
inherent phase synchronization is population pulsation within the active medium.
\end{abstract}

\vspace*{4mm}
\vspace*{4mm}




Synchronization of interacting  oscillators is encountered
in many physical, chemical, and biological systems. 
Recently, considerable research has been devoted to understanding
why and under what conditions chaotic synchronization is 
possible~\cite{BOC02}.
In chaotic synchronization, two (or more) chaotic oscillators adjust a 
given property of  their motion to a common behavior,
as a result of  coupling. Depending
on the property under consideration, different types of chaotic synchronization
were recognized; see ~\cite{BOC02} and references therein.
Chaotic phase synchronization, which is the subject of this paper,  occurs 
when  the phase difference of interacting chaotic oscillators remains 
bounded within the range of $2\pi$ for all time (phase locking)~\cite{ROS96}. 

There are different approaches to achieving chaotic synchronization.
A typical setup involves initially independent
oscillators, where at least one is chaotic. Then,
these oscillators (e.g. chaotic lasers)  are made to interact with each other, 
and the focus of existing
studies is on either an unidirectional or a bidirectional  coupling 
scheme~\cite{BOC02,ROS96,BOCprl02,BRE02,MAC03,OSI03,ROY94,PEI02,LEY03}.
Another approach uses external modulation of a system 
of interacting oscillators to force  chaotic 
synchronization, e.g. as in an experiment involving
a modulated three-mode solid state laser~\cite{OTS02}.
Less understood and perhaps most interesting is inherent chaotic synchronization
where the chaos and chaotic synchronization 
arise entirely from instabilities induced by the mutual coupling between the 
oscillators, without requiring chaos in  uncoupled oscillators. 
Inherent chaotic synchronization  was detected experimentally and
studied numerically for lasers coupled at a distance~\cite{HEI01,DES01,MIR01,MUL04}.
However, the underlying physical mechanism remains unclear. 
Moreover, in all these studies,  the chaotic-synchronized  oscillators appear 
to involve a single chaotic attractor~\cite{BOC02,ROS96,BOCprl02,BRE02,MAC03,OSI03,ROY94,PEI02,LEY03,OTS02,HEI01,DES01,MIR01,MUL04}.


This paper investigates inherent chaotic phase synchronization within the scope of 
strongly-competing oscillators.
Considering two composite-cavity modes coupled via  class-B active medium~\cite{WIEa04}
we are able to definitely identify the physical  mechanism enabling
phase synchronization to be an intrinsic behavior of this system.
Furthermore, we discovered that the chaotic phase synchronization
in strongly-competing oscillators 
exhibits bistability, a phenomenon that has not been reported previously.
Also, we found that, in contrast to chaotic phase synchronization
of oscillators that are independently chaotic,
transition to inherent chaotic phase synchronization is not clear-cut.

We consider a double-cavity laser, where cavity $A$ of length $L$ 
is coupled via a common mirror of transmission $T$
to cavity $B$ of length $L+dL$. There are two approaches to model such a system~\cite{WIEa04}.
(i) In the more phenomenological one,  lasers are treated as individual oscillators 
and the coupling is introduced via  {\it ad hoc} terms in the equations 
of motion for uncoupled lasers.
(ii) In an alternate approach, called composite-cavity mode approach, the entire coupled-laser 
structure is treated as a single system. The lasing field  is decomposed in terms of the eigenmodes
extending over both cavities rather then the fields of  individual cavities. 
In this picture, composite-cavity modes (CCMs), rather than individual lasers,
are the interacting  oscillators.
The slowly varying dimensionless electric field amplitudes $E_n$, and optical phases $\psi_n$
associated with the $n-$th CCM evolve accordingly to~\cite{WIEa04,WIEoc04}
\begin{eqnarray}
\label{eq:a}
\dot{{E}}_n&=&- \gamma {E}_n + C_{nn} \gamma \times
\sum_{k}
\left[ C_{kn}^{A}(1+\beta {N}^A)+ C_{kn}^{B}(1+\beta {N}^B)\right]\cos(\psi_{kn}) 
\,{E}_k,\\
\label{eq:b}
\dot{\psi}_n&=& {\Omega}_n + C_{nn}\gamma \times
\sum_{k}
\left[C_{kn}^{A}(1+\beta {N}^A)+ C_{kn}^{B}(1+\beta {N}^B)\right]\sin(\psi_{kn})
\frac{{E}_k}{{E}_n},
\end{eqnarray}
where $\gamma$ is the ratio between photon and population decay rates, 
$\beta$ is the dimensionless
gain coefficient,  $\Omega_n$ is the dimensionless passive 
CCM frequency, and $\psi_{kn}=\psi_k-\psi_n$~\cite{WIEoc04}. 
In class-B lasers, the active-medium polarization decays much faster then
the population and electric field. Then, the evolution of the dimensionless population $N$ 
is governed by~\cite{WIEoc04}
\begin{figure}[t]
\begin{center}

\begin{picture}(0,80)(35,0)
\put(0,3){ \epsfxsize=8cm \epsfbox{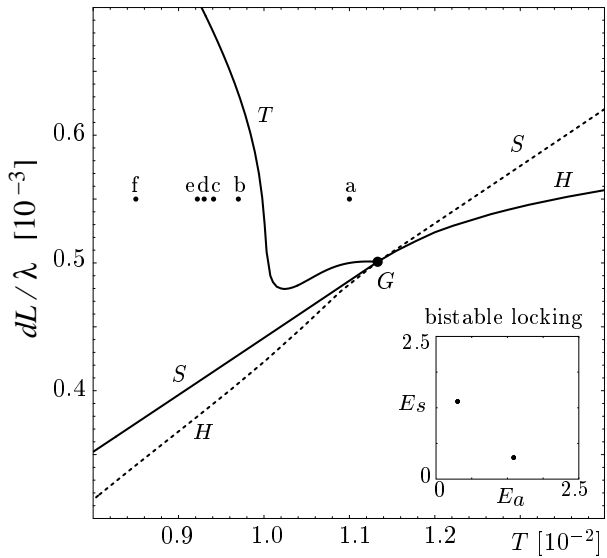} }
\end{picture}

\caption{
Bifurcation diagram near saddle-node-Hopf point $G$.
Inset shows bistability inside the locking region. The dots a-f
denote parameters for the panels in Fig.~\ref{fig:2}. 
}
\label{fig:1}
\end{center}
\end{figure}
\noindent
\begin{eqnarray}
 \label{eq:c}
\dot{{N}}_{A(B)} &=&{\Lambda}_{A(B)}-({N}_{A(B)}+1)+ 
-\sum_{m,n}C_{nm}^{A(B)}\,(1+\beta {N}^{A(B)}) \cos(\psi _{nm})\,{E}_{m}{E}_{n},
\end{eqnarray}
where $\Lambda_{A(B)}$ is the dimensionless excitation rate in cavity 
$A(B)$.
The $T$ and $dL$ dependent $C_{nn}^{A(B)}$ describes the 
overlap 
of the $n$-th CCM with the active gain medium in 
cavity $A(B)$, and  the nonlinearities due to  optical coupling 
between the cavities~\cite{WIEoc04}.
We consider two CCMs, called symmetric ($n=s$)
and antisymmetric ($n=a$) to refer to the relative optical phases between the fields in 
individual resonators~\cite{CHO86}.
For the calculations we use
$\gamma=2\times10^{11}$ s$^{-1}/(2\times10^{9}$ s$^{-1})=100$, $\beta=5.41$,
$\Lambda=2$,
and $L=280\;\mu$m~\cite{WIEoc04}.

The two sources of coupling (cross-saturation) 
between CCMs $u_n$ are
spatial and spectral hole burning.
In spatial hole burning, proportional to 
\begin{eqnarray}
\label{eq:modAB}
&&{\cal C}_{nm}=\sum_{i=A,B} C_{nn}^{i}  C_{mm}^{i}= \\
&&\frac{1}{L}
\bigg[
\int_{-L}^{0}dz\; u_n^2(z,T,L,dL)\int_{-L}^{0}dz'\; u_m^2(z',T,L,dL)+
\int_{0}^{L+dL}dz\; u_n^2(z,T,L,dL)\int_{0}^{L+dL}dz'\; u_m^2(z',T,L,dL)
\bigg],\nonumber
\end{eqnarray}
competition arises because both CCMs deplete population at the 
same locations in the active medium~\cite{WIEa04}.
In spectral hole burning, the electric field of one CCM
saturates population at the frequency of the other CCM.
One contribution to spectral hole burning comes from
population pulsation, a result of nonlinear composite-mode 
interaction where the active-medium population acquires an oscillation at the intermode 
frequency [$n\neq m$ terms in Eq.~(\ref{eq:c})]. 
This oscillation interacts with the $n-$th  CCM to modify the 
active-medium polarization at the frequency of the $m-$th CCM. 
Consequently, more competition arises due to additional cross-saturation 
of the $m-$th CCM. 
Under appropriate conditions, the additional cross-saturation 
leads to strong competition 
resulting in  bistability between stable stationary points~\cite{CHO86,WIEoc04}.

Coupling between CCMs is adjusted with $T$ and $dL$.
For $T=0$ and $dL\neq 0$, ${\cal C}_{sa}=0$ and the two CCMs are uncoupled:
each CCM becomes a mode of a different individual laser~\cite{CHO86}.
Recall, that we study inherent chaotic synchronization 
and the two CCMs are not chaotic at zero coupling.
For $|dL|>0$, coupling increases 
with $T$~\cite{WIEoc04}. As a result, the two 
interacting CCMs turn  chaotic and  exhibit transition to chaotic 
phase synchronization.

Applying bifurcation continuation techniques~\cite{AUTO}
to Eqs.~(\ref{eq:a}-\ref{eq:c}), we calculated
saddle-node $S$, Hopf $H$, and torus $T$ bifurcation curves
in the parameter space $(T,dL/\lambda)$ [Fig.~\ref{fig:1}]. 
Curves $S$ and $H$ are tangent at 
saddle-node-Hopf points $G$ where they change their type. 
Supercritical bifurcations of attractors are plotted as
solid curves and subcritical bifurcations of unstable objects are 
plotted as dashed curves.
Inside the lockband, which extends below the solid parts of $S$ and $H$, 
composite modes are phase locked to 
operate at constant intensity and the same optical frequency.
Moreover, there exist two stable  stationary points 
in the $\{E_a,E_s,\psi_{sa},N_A,N_B\}$  phase space.
Depending on initial conditions, the system of coupled CCMs settles
through strong competition to oscillate at an optical  frequency that is 
\begin{figure}[t]
\begin{center}

\begin{picture}(88,190)(0,0)
\put(0,0){ \epsfxsize=8.5cm \epsfbox{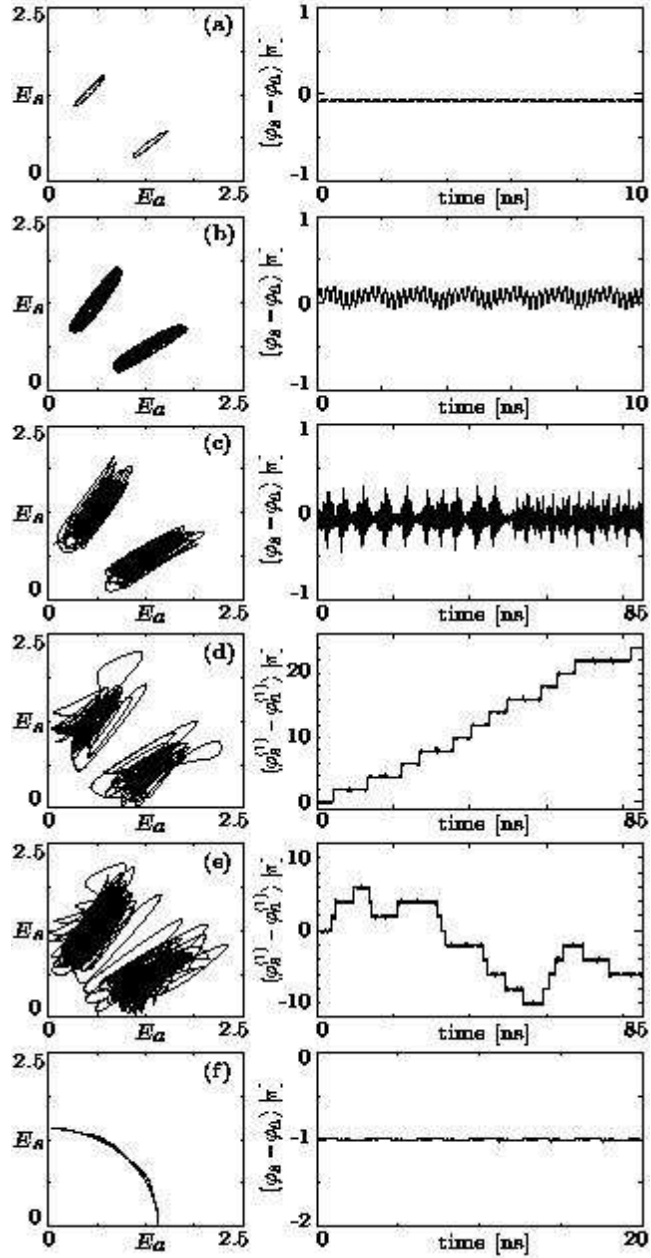} }
\end{picture}

\caption{
(first column) Phase portraits and (second column)
time evolution of the analytic-phase difference for 
the transition at $dL/\lambda=0.55\times10^{-3}$. From (a) 
to (f) $T\; [10^{-2}] =$1.1, 0.97, 0.94, 0.93, 0.924, 
and 0.85. Refer to the dots a-f in Fig.~\ref{fig:1}.
The time evolution of the analytic-phase difference is shown
for the lower of the two coexisting attractors.
}
\label{fig:2}
\end{center}
\end{figure}
\noindent
near the frequency 
of either  symmetric or antisymmetric CCM.
Each bifurcation curve in Fig.~\ref{fig:1} denotes bifurcations
of the two  attractors~\cite{WIEoc04}.

Outside the lockband, optical-phase locking to a single optical frequency is 
lost and  the modal intensities oscillate. Even when they are irregular, 
these oscillations may still exhibit certain types of synchronization 
when described with  appropriate variables.  
Here, the appropriate quantity is the  analytic phase $\varphi_{k}$ 
of a real signal $x_{k}(t)=E^2_{k}(t)-\langle E^2_{k}(t) \rangle$,
that  is defined through 
$x_{k}(t)+i\tilde{x}_{k}(t)=\sqrt{x_{k}^2(t)+\tilde{x}_{k}^2(t)}\exp[i\varphi_{k}(t)]$,
where  $\langle E^2_{k}(t) \rangle=\lim_{\,T \to \infty}\int_0^T \left[E_k^2(t)/T\right] dt$,
${\tilde x}_{k}$
is the Hilbert transform of $x_{k}$~\cite{GAB46}, and  $k=a,s$. 
Analytic phase $\varphi_{k}$ is well defined if the trajectory 
in the complex plane $(x_k,{\tilde x_k})$
has one center of rotation. In the case of multiple centers of rotation, 
the signal  needs to be
decomposed into 
\begin{figure}[t]
\begin{center}

\begin{picture}(88,40)(0,0)
\put(5,3){ \epsfxsize=8cm \epsfbox{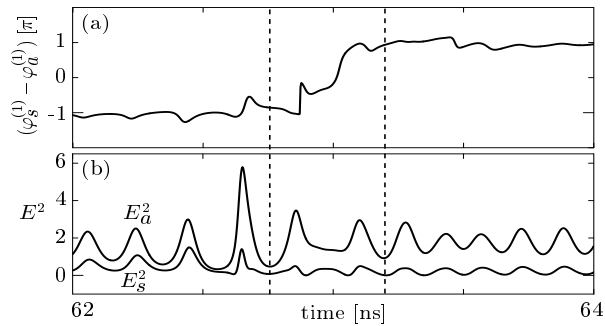} }
\end{picture}

\caption{(a) An example of a $2\pi$ jump in the analytic-phase difference 
[compare with Fig.~\ref{fig:2} (d)]
and (b) the corresponding chaotic oscillations of
the coupled composite-cavity modes.
}
\label{fig:3}
\end{center}
\end{figure}
\noindent
intrinsic modes $x_k(t)=\sum_j{ x_k^{(j)}(t)}$ such that the trajectory
in the plane $(x_k^{(j)},{\tilde x_k^{(j)}})$ has one center of 
rotation~\cite{YAL97}. In this paper, we use $\varphi_{k}^{(1)}$ whenever chaotic attractor 
has multiple centers of rotation. The analytic phase $\varphi_k$ of the modal intensity 
is distinctly different from the optical phase  $\psi_k$.

When the solid part of $H$ is crossed from below to above in Fig.~\ref{fig:1}, 
locking of optical phases is lost. Each of the two coexisting
stationary points turns unstable and gives rise to a stable periodic orbit.
The relaxation oscillation, which is a characteristic of a class-B laser, 
become undamped. 
Now, both CCMs selfpulsate but their
analytic phases are locked [Fig.~\ref{fig:2} (a)].
Periodic oscillation can undergo further instabilities.
Torus bifurcation curve $T$ emerging from $G$ marks 
the transition to quasiperiodic oscillation that involves two frequencies: 
the relaxation oscillation  frequency
and the CCM beatnote. 
Dynamics on the two tori also show analytic-phase synchronization [Fig.~\ref{fig:2} (b)].
Decreasing $T$ further leads to the break-up of coexisting tori into
chaotic attractors. Despite chaotic oscillations 
in both CCMs, their analytic-phase difference  remains within 
a range of $2\pi$ for all time [Fig.~\ref{fig:2} (c)]. 
Inherent chaotic phase synchronization is present simultaneously at two
different chaotic attractors.
Upon further decrease in the coupling, analytic-phase synchronization is lost [Fig.~\ref{fig:2} (d)].
The analytic-phase difference for the 
lower (upper) chaotic attractor 
increases (decreases) in time exhibiting  $2\pi$ jumps (Fig.~\ref{fig:3}).
Next, the two chaotic attractors merge into one chaotic 
attractor [Fig.~\ref{fig:2} (e)]. The analytic phases remain unlocked but 
the direction of $2\pi$ jumps alternates
as a result of transitions between the two formerly bistable  chaotic attractors.

\begin{figure}[t]
\begin{center}

\begin{picture}(120,80)(0,0)
\put(0,3){ \epsfxsize=12cm \epsfbox{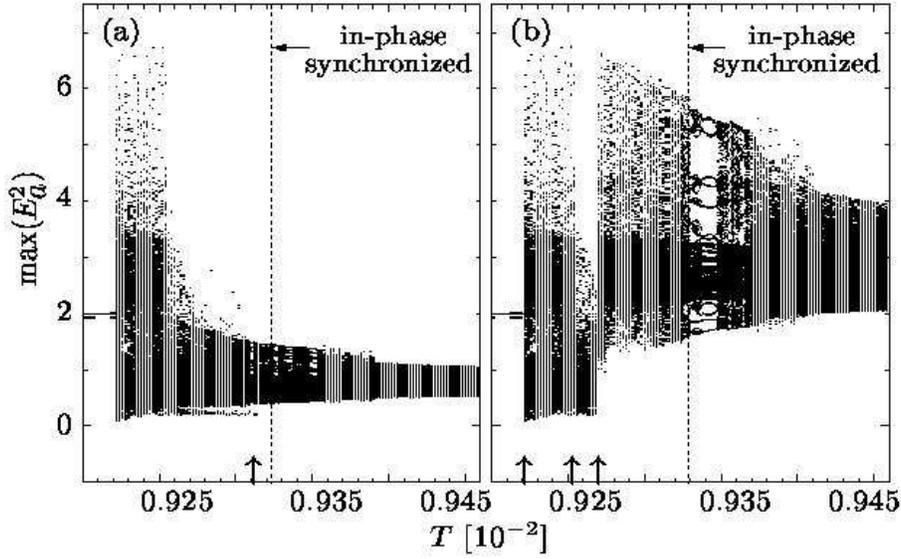} }
\end{picture}

\caption{Bifurcations of the (a) upper and (b) lower chaotic attractor
with decreasing $T$, leading to a loss of phase synchronization 
and subsequently to a single chaotic attractor.
}
\label{fig:4}
\end{center}
\end{figure}

\begin{figure}[t]
\begin{center}

\begin{picture}(120,100)(0,0)
\put(0,0){ \epsfxsize=12cm \epsfbox{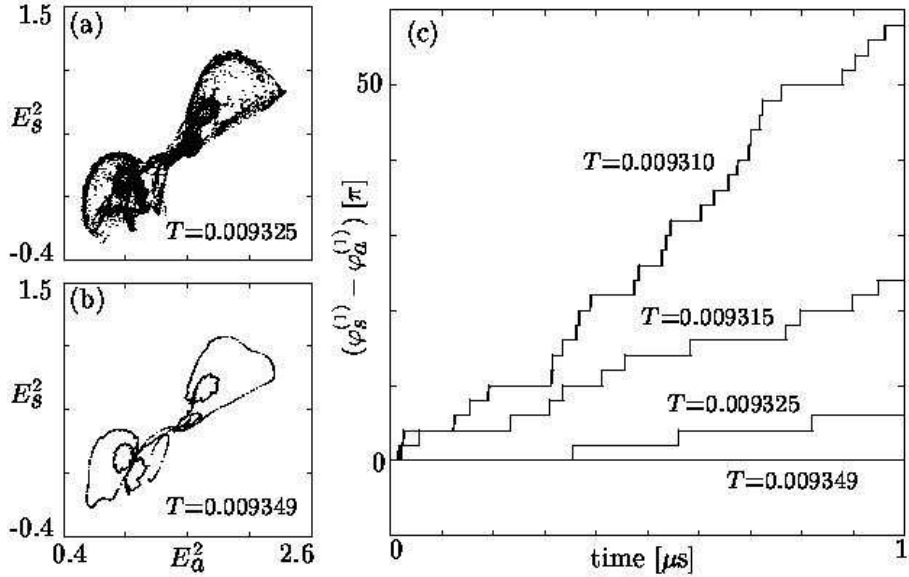} }
\end{picture}

\caption{
(a-b) Change in the internal structure of the lower chaotic attractor 
due to transition through an Arnold tongue
shown as a Poincar\'e section defined by $\{N_A=0.025\}$ 
[compare with Fig.~\ref{fig:4}(b)], 
and (c) the resulting loss of chaotic phase synchronization.
}
\label{fig:5}
\end{center}
\end{figure}

To see how the changes in analytic-phase synchronization of coupled chaotic 
CCMs come about,  we plot in Fig.~\ref{fig:4} the maxima of $E_a^2$ versus $T$.
The simulation starts at higher $T$ with two chaotic attractors, the lower and the upper. 
As $T$ decreases, changes in the chaotic dynamics appear, including windows of periodicity 
called Arnold tongues.
We find that analytic-phase synchronization is lost after transition 
through an Arnold tongue. At the right hand side of the widest Arnold tongue in Fig.~\ref{fig:4}(b),
periodic orbit of high period is born and replaces the chaotic attractor.
After some bifurcations, this orbit disappears to give place to a new chaotic attractor.
However,  the  chaotic attractor that appears at the left  hand side of this  Arnold tongue
[Fig.~\ref{fig:5}(a)] is  significantly  different 
from the chaotic attractor that disappears at the right hand side of this  Arnold tongue 
[Fig.~\ref{fig:5}(b)]. The change(s) in the structure of the chaotic attractor
(possibly due to homoclinic or heteroclinic tangency between stable and 
unstable manifolds of some saddle orbits) result in phase 
desynchronization. Analytic-phase difference for
chaotic attractor  at the left hand side of the Arnold tongue 
shows $2\pi$ phase jumps that become more frequent with decreasing $T$ [Fig.~\ref{fig:5}(c)]. 
Similar changes take place in the upper chaotic attractor.
Although the detailed bifurcation 
scenario inside the (narrow) Arnold tongue in Fig.~\ref{fig:4}(a) is different, the resulting 
effect is the same: phase-desynchronized chaotic attractor emerges. 
It is important to note that within desynchronized chaos we find  Arnold tongues with 
dynamics that still show analytic-phase synchronization 
[arrow at the bottom of Fig~\ref{fig:4}(a)]. Therefore, transition 
to  chaotic phase synchronization of independently stable oscillators is 
not as clear-cut as  for oscillators that are independently chaotic ~\cite{ROS96}.

As $T$ decreases further, the lower chaotic 
attractor hits the basin boundary 
that separates the two chaotic 
attractors, and is destroyed
[right arrow at the bottom of Fig~\ref{fig:4}(b)]. 
The trajectory settles to the upper chaotic attractor
which then   expands onto the remnants of the lower chaotic attractors
[middle arrow at the bottom of Fig~\ref{fig:4} (b)]. These two crises lead to a single chaotic attractor
which is composed of the two formerly bistable attractors.
Eventually, the chaotic attractor from Fig.~\ref{fig:2} (e) 
hits the boundary that separates it from a periodic orbit [left arrow at the bottom of Fig~\ref{fig:4}(b)]
and the trajectory settles to the  periodic orbit with 
anti-phase dynamics [Fig.~\ref{fig:2} (f)]. 
This orbit originates  at the solid part of $S$ and has a different origin than the periodic 
orbits in Fig.~\ref{fig:2} (a). 
It arises due to unlocking of optical phases of the two CCMs~\cite{WIEa04,WIEoc04}.
Upon increasing $T$, this anti-phase oscillation coexists 
with   in-phase oscillations, until it disappears in a saddle-node-of-periodic-orbit bifurcation
at $T\sim 1.1\times10^{-2}$.

Different physical mechanisms are responsible for  in-phase and out-of-phase
synchronization.
On the one hand, 
\noindent
\noindent
when  composite-mode detuning is too large for the 
optical phases to  lock, but 
close to  the relaxation oscillation frequency
($T\gtrsim 1.1\times 10^{-2}$ and outside lockband),
population pulsation provides an inherent source of modulation that can
excite self-sustained relaxation oscillation and force in-phase 
dynamics of modal intensities.
On the other hand, when CCM  frequencies are close 
enough ($T\lesssim 1.1\times 10^{-2}$ and outside lockband)
the optical-phase locking terms become significant. 
Then, the system of two coupled CCMs  alternates between 
the two states of nearly 
single-CCM operation, resulting in anti-phase dynamics.
For  intermediate conditions, in-phase and anti-phase dynamics can coexist.
Strong population pulsation effects are crucial for the inherent
chaotic phase synchronization to occur. If they are neglected,
neither bistability nor the instabilities leading to  
synchronization are present.

In conclusion, we extended the usual analysis of chaotic phase synchronization
to strongly-competing oscillators, where system dynamics are more diversified
and less understood. In addition to previously reported results, the present analysis  shows that
chaotic phase 
synchronization can occur in coupled oscillators that are independently stable.
This inherent chaotic phase synchronization results solely from the nonlinearities 
associated with strong mode competition within a saturable active medium.
In the presence of population pulsation, strong competition causes the phenomenon to appear 
simultaneously at two different chaotic attractors, giving rise to bistable 
chaotic-phase-synchronized solutions.
In contrast to phase synchronization of independently chaotic oscillators, 
transition to (inherent) phase synchronization of independently stable oscillators 
is not clear-cut: windows of phase-synchronized dynamics are found within 
phase-desynchronized chaos.


\vspace*{5mm}
This work is partially funded by the US Department of Energy 
under contract DE--AC04--94AL8500. WC acknowledges support from the 
Research Award of the Alexander von Humboldt Foundation.




\vspace*{0mm}

\begin{thebibliography}{}


\bibitem{BOC02} S. Boccaletti et al., 
{\em  Phys. Rep.}  {\bf 366 } (2002) 1.

\bibitem{ROS96} M.G. Rosenblum, A.S. Pikovsky, and J. K\"urths, 
{\em Phys. Rev. Lett.}  {\bf 76} (1996) 1804.


\bibitem{BOCprl02}S. Boccaletti et al., 
{\em Phys. Rev. Lett.}  {\bf 89} (2002) 194101.

\bibitem{BRE02}R. Breban and E. Ott,
{\em Phys. Rev. E.}  {\bf 65} (2002) 056219.

\bibitem{MAC03}R. McAllister et al., 
{\em Phys. Rev. E}  {\bf 67} (2003) 015202(R).

\bibitem{OSI03} G.V. Osipov et al., 
{\em Phys. Rev. Lett.} {\bf 91} (2003) 024101.

\bibitem{ROY94} R. Roy and K.S. Thornburg Jr.,
{\em Phys. Rev. Lett.}  {\bf 72} (1994) 2009.

\bibitem{PEI02} M. Peil, et al.
{\em Phys. Rev. Lett.}  {\bf 88} (2002) 174101.

\bibitem{LEY03} I. Leyva et al., 
{\em Phys. Rev. E} {\bf 68} (2003) 066209.


\bibitem{OTS02} K. Otsuka et al., 
{\em Chaos}  {\bf 12} (2002) 678.

\bibitem{HEI01}T. Heil, I. Fischer, and W. Els\"a\ss er, 
{\em  Phys. Rev. Lett.}  {\bf 86} (2001) 795.

\bibitem{DES01} D.J. DeShazer et al., 
{\em  Phys. Rev. Lett.}  {\bf 87} (2001) 044101.

\bibitem{MIR01} C.R. Mirasso et al.,
{\em  Phys. Rev. A}  {\bf 65} (2001) 013805.

\bibitem{MUL04} J. Mulet et al., 
{\em J. Opt. B: Quant. Semiclass. Opt.}  {\bf 6} (2004) 97.







\bibitem{WIEa04} S. Wieczorek and  W.W. Chow,
{\em Phys. Rev. A} {\bf 69} (2004) 033811.

\bibitem{CHO86} W.W.Chow,
{\em IEEE Jour. of Quant. Electron.}  {\bf QE-22} (1986) 1174.

\bibitem{WIEoc04} S. Wieczorek and  W.W. Chow,
{\em Opt. Comm.}, to appear.


\bibitem{AUTO}E. Doedel et al., 
{\em http://sourceforge.net/ projects/auto2000/}.

\bibitem{GAB46} D. Gabor,
J. IEEE (London)  {\bf 93} (1946) 429.

\bibitem{YAL97} T. Yal\c cinkaya and Y-C. Lai,
{\em Phys. Rev. Lett.} {\bf 79} (1997) 3885.




\end{thebibliography}
\end{document}